\documentclass[sigconf,balance=false]{acmart}
\usepackage{popets}
\usepackage{subcaption}
\usepackage{xspace}
\usepackage{lipsum}
\usepackage{breakcites}
\usepackage{balance}

\setcopyright{popets}
\copyrightyear{YYYY}

\acmYear{YYYY}
\acmVolume{YYYY}
\acmNumber{X}
\acmDOI{XXXXXXX.XXXXXXX}
\acmISBN{}
\acmConference{Proceedings on Privacy Enhancing Technologies}
\newcommand{\Panon}{\emph{Prob(2-anon)}\xspace}
\begin{document}

\title[On the Robustness of Topics API to a Re-Identification Attack]{On the Robustness of Topics API to a Re-Identification Attack}

\author{Nikhil Jha}
\affiliation{%
  \institution{Politecnico di Torino}
  \city{Torino}
  \state{}
  \country{Italy}}
\email{nikhil.jha@polito.it}

\author{Martino Trevisan}
\affiliation{%
  \institution{Universit\`a degli Studi di Trieste}
  \city{Trieste}
  \state{}
  \country{Italy}}
\email{martino.trevisan@dia.units.it}

\author{Emilio Leonardi}
\affiliation{%
  \institution{Politecnico di Torino}
  \city{Torino}
  \state{}
  \country{Italy}}
\email{emilio.leonardi@polito.it}

\author{Marco Mellia}
\affiliation{%
  \institution{Politecnico di Torino}
  \city{Torino}
  \state{}
  \country{Italy}}
\email{marco.mellia@polito.it}

\renewcommand{\shortauthors}{Jha et al.}

\begin{abstract}
Web tracking through third-party cookies is considered a threat to users' privacy and is supposed to be abandoned in the near future. Recently, Google proposed the Topics API framework as a privacy-friendly alternative for behavioural advertising. Using this approach, the browser builds a user profile based on navigation history, which advertisers can access. The Topics API has the possibility of becoming the new standard for behavioural advertising, thus it is necessary to fully understand its operation and find possible limitations.

This paper evaluates the robustness of the Topics API to a re-identification attack where an attacker reconstructs the user profile by accumulating user's exposed topics over time to later re-identify the same user on a different website. Using real traffic traces and realistic population models, we find that the Topics API mitigates but cannot prevent re-identification to take place, as there is a sizeable chance that a user's profile is unique within a website's audience. Consequently, the probability of correct re-identification can reach $15-17\%$, considering a pool of 1,000 users. We offer the code and data we use in this work to stimulate further studies and the tuning of the Topic API parameters.
\end{abstract}


\maketitle

\section{Introduction}
\label{sec:intro}

In the current web ecosystem, targeted or behavioural advertising lets providers monetize their content, by collecting and processing personal data to build accurate user profiles. Among the techniques, web tracking is the most widespread technology~\cite{mayer2012third,metwalley2015online,englehardt2016online}. It heavily leverages third-party cookies, that allow tracking platforms to follow the same user on different websites. The mechanism can be summarized as follows: when a user visits a website, a tracker installs a third-party \emph{profiling} cookie on the user's client. This cookie contains a unique identifier that lets the tracker identify the user on subsequent visits. When the user visits a second website that embeds the same tracker, the cookie is sent to the tracker, as specifications mandate that cookies are handled on a per-domain basis. As such, the tracker learns that the same user has visited the two websites. Using this mechanism, trackers gather information on users and build profiles describing their interests. Profiles are offered to advertisers so that they can customize the content of the displayed ads. In some cases, tracking platforms employ more sophisticated and privacy-intrusive techniques such as browser fingerprinting or ID synchronization~\cite{rizzo2021unveiling,papadogiannakis2021}. 

This massive data collection has created tension between users and the ads ecosystem~\cite{sipior2011online,mayer2012third,estrada2017online}. Some browsers, such as Mozilla Firefox and Apple Safari, have already started battling third-party cookies. Leading researchers and industries are studying new paradigms that are more respectful of users' privacy.

These new proposals have one common feature: the replacement of third-party cookies and tracking with other techniques that let the user control and limit the amount of disclosed personal information. First, Google proposed the Federated Learning of Cohorts (FLoC)~\cite{ravichandran2021evaluation}. In FLoC, users are clustered in cohorts according to their interests, computed by each one's browser based on the user's recent activity. In the proponents' intentions, this solution should have prevented tracking, as every user was ``hidden'' in his/her cohort. However, the approach has been criticised~\cite{rescorla2021technical}. The main issue is that while a user could hide inside a cohort for a short period of time, the sequence of cohorts they belonged to across time could work as an identifier, increasingly unique. Eventually, Google replaced it with a new proposal called \emph{Topics API}. With the Topics API, the browser is in charge to build the user's profile based on the navigation history. Websites can ask for a privacy-preserving version of such profiles to serve targeted advertisements or services. Among the mechanisms in place, the Topic API returns at most one topic per epoch (a week), and randomly replaces 5\% of actual topics with random ones.

The Topics API framework has the potential to become the new standard for behavioural advertising and replace the current conflicting web tracking system based on third-party cookies. In the first quarter of 2024, Google will deprecate the use of cookies for 1\% of Chrome users, to ``support developers in conducting real world experiments that assess the readiness and effectiveness of their products without third-party cookies''\footnote{\url{https://privacysandbox.com/news/the-next-stages-of-privacy-sandbox-general-availability}, accessed on \today}. It is thus urgent to fully understand the operation of the Topics API, and independent researchers must verify the robustness of such an approach as done by Mozilla and Google~\cite{epasto2022measures,thomson2023privacy}.

In this paper, we provide an independent evaluation of the Topics API. Using a data-driven approach, we build realistic population models that we use to quantify the feasibility of a re-identification attack: We assume that the attacker i) exploits the Topic API to reconstruct the victim's profile by accumulating her/his topics over epochs, and ii) tries to re-identify the victim on the audience of a second website -- as studied by Epasto et al.~\cite{epasto2022measures}. If successful, such an attack would tamper with the abandonment of third-party cookies, allowing platforms to still track users across websites.
We face the problem by mapping it to the probability that a user is $k$-anonymous among the website audience, i.e., that there are $k-1$ other users with the same reconstructed profile. Generalising the attack sketched by Thomson~\cite{thomson2023privacy},
we propose a robust denoising algorithm  that aims to filter the random topics introduced by the Topics API.

We contribute to three main results:
\begin{enumerate}
    \item  We show that the introduction of Topics API algorithm mitigates but cannot prevent re-identification. Depending on the website's audience size (e.g., 100,000 visitors) and population heterogeneity, a sizeable fraction (e.g., 40\%) of users would still let the attacker reconstruct a denoised and unique profile that allows re-identification if matched on a second population.
    \item We demonstrate the replacement of actual topics with random ones is key to limiting the attack. Yet, the denoising algorithm is very efficient in removing random topics from the reconstructed profiles the attacker builds.
    \item We show that in practice the probability of correctly re-identify a user in a pool of 1,000 can top 15-17\%, with false positives being negligible (less than 0.2\%). However, it is also important to consider that such probabilities are a function of the attacker's observation period and that many weeks may be needed to carry out the attack in practice.
    
\end{enumerate}
Our study highlights the need for continued research and development of privacy-preserving advertising techniques to ensure that user privacy is respected in the digital age. To foster research in this field, we release the code and data to replicate and extend our experiments.\footnote{The code is available at \url{https://github.com/nikhiljha95/topics-api-simulator}.}

The remainder of the paper is organized as follows: Section~\ref{sec:metho} formalizes Topics API operation and the threat model. In Section~\ref{sec:dataset} and~\ref{sec:models}, we describe the dataset and models to generate synthetic populations we use to run simulations, respectively. Section~\ref{sec:results} illustrates the results in terms of $k$-anonymity, while Section~\ref{sec:attack} explores the effectiveness of a re-identification attack. Section~\ref{sec:related} summarizes related work, and, finally, Section~\ref{sec:conclu} discusses our findings and concludes the paper.
\section{The Topics API and the Threat Model}
\label{sec:metho}

In this section, we describe how the Topics API operates for creating a profile from the user's browsing history. Then, we describe our threat model -- i.e., the possibility that an attacker links two profiles referring to the same user as they are uniquely identifiable within a given population.

We consider a browser that a user employs to navigate the Internet.\footnote{We intentionally confuse the term \textit{user} and \textit{browser} to identify the person and the application they use to navigate the Internet.}  We assume time is divided into epochs of duration $\Delta T$ (one week in the current proposed Topics API operation). During each epoch $e$, the browser collects and counts the number of visits to each website and forms a \emph{bag of websites}  $\mathcal{B}_{u,e}$ for the user $u$. It keeps track only of the website hostnames the user \textit{intentionally} visited, e.g., by typing its URL, or by clicking on a link in a web page or other applications. Formally, given a user $u$ and the epoch $e$, let $\mathcal{B}_{u,e} = \{ (w_1, f_{1,u,e}),  (w_2, f_{2,u,e}), \ldots, (w_n, f_{n,u,e}) \}$, where $\{w_{i}\}$ represent the visited websites and $f_{i,u,e}$ the number of times $u$ visited $w_i$ during epoch $e$.

\begin{table}[]
\centering
\caption{Main terminology to model Topics API algorithm and threat model.}
\label{tab:terminology}
\begin{tabular}{@{}ll@{}}
\toprule
\textbf{Symbol}         & \textbf{Definition}                                                 \\ \midrule
$ n_{topic} $           & Number of topics in the taxonomy                                    \\
$ E $                   & Number of past epochs included in the profile                       \\
$ p $                   & Probability a random topic to replace a real topic                  \\
$ N $                   & Epochs of observation by the attacker                               \\
$ U $                   & User population set                                                 \\
$ \lambda_{u,t} $       & Rate of visit by user $u$ to topic $t$                              \\
$ \mathcal{B}_{u,e} $   & Bag of visited \emph{websites} by user $u$ at epoch $e$             \\
$ \mathcal{T}_{u,e}$    & Bag of visited \emph{topics} by user $u$ at epoch $e$               \\
$ \mathcal{P}_{u,e} $   & Profile for the user $u$ at epoch $e$                               \\
$ \mathcal{P}_{u,e,w} $ & Exposed Profile to website $w$ for user $u$ at epoch $e$            \\
$ \mathcal{G}_{u,N,w} $ & Global Reconstructed Profile by $w$ after $N$ epochs                \\
$ \mathcal{R}_{u,N,w} $ & Denoised Reconstructed Profile by $w$ after $N$ epochs               \\
\bottomrule
\end{tabular}
\end{table}

\subsection{The Topics API profile construction}
\label{sec:topics-api}

The Topics API algorithm operates in the browser and processes the history of $\mathcal{B}_{u,e}$ over the past $E$ epochs to create a corresponding \textit{Exposed Profile} $\mathcal{P}_{u,e,w}$ for the user $u$, epoch $e$ and each specific website $w$ the user visits during the current epoch. In fact, the browser builds a separate \textit{Exposed Profile} for each visited website $w$ to mitigate re-identification attacks. We base the following description on the public documentation of the Topics API available online.\footnote{\url{https://developer.chrome.com/docs/privacy-sandbox/topics/}, accessed on \today} The operation of the Topics API has the following steps.

\vspace{2mm}
\noindent
\emph{\bf Step~1 - From websites to topics}: For each of the websites $w_i\in \mathcal{B}_{u,e}$, the browser extracts a corresponding \textit{topic} $t_i$. To this end, the browser uses a Machine Learning (ML) classifier model that returns the topic of a website given the characters and strings that compose the website hostname. 

At this step, each browsing history $\mathcal{B}_{u,e}$ is transformed into a \textit{topic history} $\mathcal{T}_{u,e} = \{ (t_1, f'_{1,u,e}),  (t_2, f'_{2,u,e}), \ldots, (t_m, f'_{m,u,e}) \}$ where $t_{i}$ represents the topic the model outputs, and $f'_{i,u,e}$ counts its total occurrences. Each website is mapped to a topic and the original frequencies $f_{i,u,e}$ are summed by topics into $f'_{j,u,e}$. There are $n_{topic}$ which form a taxonomy of possible interests the users have. Such taxonomy will include between a few hundred and a few thousand topics (the IAB Audience Taxonomy contains about 1,500 topics)\footnote{\url{https://iabtechlab.com/standards/audience-taxonomy/}, accessed on \today}. In our experiments, we employ the Google ML model implemented in Chrome. In its current implementation, it supports $n_{topic}=349$ topics\footnote{\url{https://github.com/patcg-individual-drafts/topics/blob/main/taxonomy_v1.md}, accessed on \today} and the model is based on a Neural Network trained by Google using a manually curated set of 10,000 domains. It leverages website hostnames only and neglects any other part of a URL.\footnote{The mapping from a website to a category is prone to inaccuracies and depends on the employed ML model. Here we do not consider the implications of such errors. See \url{https://developer.chrome.com/docs/privacy-sandbox/topics/\#classifier-model}, accessed on \today}

\vspace{2mm}
\noindent
\emph{\bf Step~2 - From Topics to Profiles}: 
Given the topic history $\mathcal{T}_{u,e}$ for user $u$ at epoch $e$, the browser selects the $z$ most frequently visited topics and stores them into the \emph{Profile history} $\mathcal{P}_{u,e}$, which will be referred as the user $u$ Profile at epoch $e$ in the following. $z$ is currently put to 5.

\vspace{2mm}
\noindent
\emph{\bf Step~3 - Per-website topic selection}: The first time the user visits the website $w$, the browser generates a \textit{Exposed Profile} $\mathcal{P}_{u,e,w}$. For each past epoch $i\in\{e-1,\ldots,e-E\}$, the browser selects at random one topic $t^*_i$ from the Profile history $\mathcal{P}_{u,i}$. $\mathcal{P}_{u,e,w}$ contains thus at most $E$ topics. To increase privacy guarantees, at each extraction, with probability $p$ the browser replaces the topic $t^*_i$ with a random topic $t_{rnd}$ uniformly selected from the global topic list. $p$ is currently suggested to be 0.05. $\mathcal{P}_{u,e,w}$ contains thus at most $E$ topics (a topic picked from $\mathcal{P}_{u,e-1}$, a topic from $\mathcal{P}_{u,e-2}$, etc.). Once generated, the Exposed Profile remains the same for the whole epoch $e$. 

\vspace{2mm}
\noindent
\emph{\bf Usage by websites}: From this point on, each time the user visits the website $w$ during the current epoch, the website $w$ may request the browser to share the current Exposed Profile $\mathcal{{P}}_{u,e,w}$ and use the returned topics to provide behavioural advertising. Notice that the Exposed Profile $\mathcal{P}_{u,e,w}$ is built only for websites intentionally (first-party) visited by the  user $u$. Any third-party service (e.g., a component embedded on the webpage of site $w$, but hosted on a different domain) will receive topics of the first-party websites $w$ it is embedded into. That is, all trackers embedded into the website $w$ receive always the Exposed Profiles $\mathcal{P}_{u,e,w}$ of $w$.

\vspace{2mm}
\noindent
\emph{\bf Periodic Profile update}: At the beginning of the epoch $e+1$, the browser computes the new Profile history  $\mathcal{P}_{u,e+1}$ and discards $\mathcal{P}_{u,e-E}$. Similarly, if and when the user visits again the website $w$, the browser creates $\mathcal{P}_{u,e+1,w}$ from $\mathcal{P}_{u,e,w}$: it  includes a new topic selected from $\mathcal{P}_{u,e+1}$ (Step 3), and removes the oldest topic, i.e., the one originally belonging to $\mathcal{P}_{u,e-E+1}$ (keeping the others). This means that a website continuously visited by a user can observe up to one new topic per epoch (and such topic may be randomly extracted).


\subsection{Threat model}
\label{sec:threat}

\begin{figure}
    \centering
    \includegraphics[width=\columnwidth]{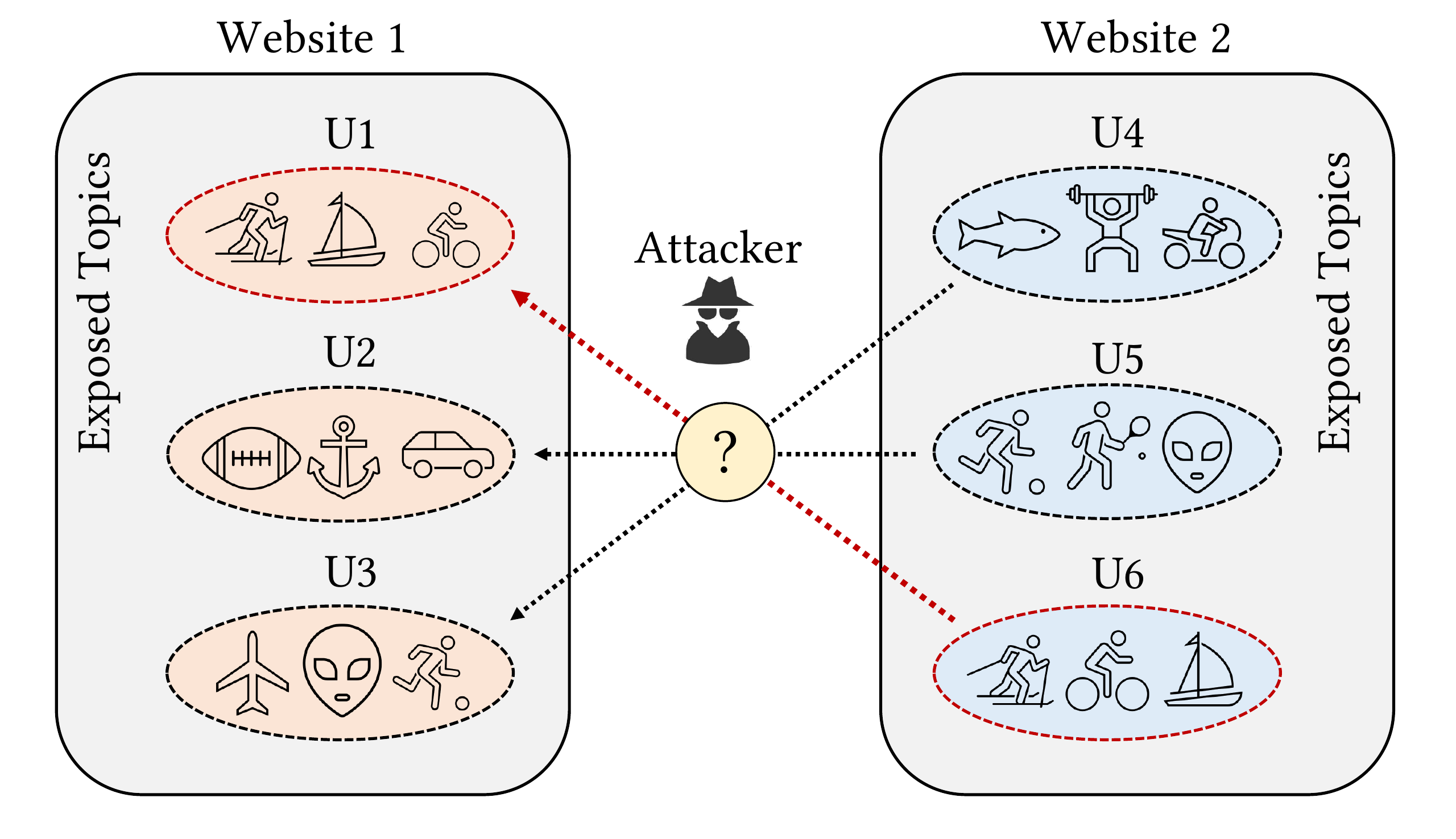}
    \caption{Threat model sketch: An attacker leverages the Exposed Profiles obtained from the Topics API to re-identify a user in the population of two websites.}
    \label{fig:threat}
\end{figure}

In this paper, we consider the threat model introduced by the same proponents of Topics API~\cite{epasto2022measures} and discussed in a technical report by Mozilla~\cite{thomson2023privacy}. In detail, we consider the risk of re-identification -- i.e., the possibility to link a \textit{Reconstructed User Profile} from an audience to a known individual; or that two websites use the Reconstructed User Profiles to match their audiences. Such possibility has already been evaluated in the literature on similar contexts~\cite{olejnik2012why,herrmann2013behavior,vassio2017users}. We sketch the second attack in Figure~\ref{fig:threat}.

\subsubsection{The re-identification threat model}  As in~\cite{epasto2022measures}, we assume a website $w$ uses first-party cookies to track a user over time so that it can reconstruct the set of topics users in its audience are interested in. Then, it matches the derived profiles with the target profile of the victim (or with all profiles of the second website audience). 

In this attack, the attacker accumulates the Exposed Profiles $\mathcal{P}_{u,e,w}$ over epochs, overcoming the limitation introduced by  Topics API to limit the Exposed Profiles to one topic per epoch, for at most $E$ epochs.
Let us assume $w$ observes its users $u\in U(w)$ for $N$ epochs (i.e., epochs in $[1,N]$). At the end of the process, for each user $u$, it builds the \textit{Global Reconstructed User Profile} as 
$\mathcal{G}_{u,N,w}= \cup_{e\in [1,N]}  \mathcal{P}_{u,e,w}$.
In the long run, the set of topics could act as an identifier string (or fingerprint)  for  user $u$, enabling the re-identification process either with the set of topics of a known user or with users from the audience $U_2$ of website $w_2$. 

Notice that this attack may be carried out by a third party too. In this case, we assume some websites $w_1$ and $w_2$ collude with a third-party service $s$. Both $w_1$ and $w_2$ embed $s$. They both share with $s$ the user identifier each time a user visits them. The third party then builds $\mathcal{G}_{u,N,w_1}$ and $\mathcal{G}_{u,N,w_2}$ autonomously so that it can match the profiles of users in both audiences.\footnote{Notice not every third-party $s$ will receive a topic. Only if $s$ observed the user visit a site $w$ about the topic in question within the past $E$ weeks, then $s$ is allowed to receive such a topic (see \url{https://github.com/patcg-individual-drafts/topics}). We ignore such limitation, i.e., we assume that the third party $s$ is pervasive enough to make this condition irrelevant because the third party is present on the most popular websites, which will enable the reception of every topic. This is the case with popular web trackers.}

\subsubsection{Random topic replacement to prevent the attack}
Being aware of such an issue, the Topics API algorithm injects random topics with probability $p$ in the Exposed Profile -- see Step 3 in the previous section. This has the benefit of making the Global Reconstructed Profile $\mathcal{G}_{u,N,w}$ both noisy (thus preventing the exact re-identification with a known victim profile) and potentially identical for all users (i.e., for $N\to\infty$, all users' Global Reconstructed Profiles would include all topics).
Notice that the injection of random topics runs separately for each website so that the Exposed Profiles on the two websites would be different.

Therefore, the attacker is somehow forced to identify and eliminate from $\mathcal{G}_{u,N,w}$
topics that are likely random, obtaining an
hopefully \textit{Denoised Reconstructed Profile} $\mathcal{R}_{u,N,w}\subseteq \mathcal{G}_{u,N,w}$ on which exact matching can be used as a reasonable criterion for identification. We present a discussion on how $\mathcal{R}_{u,N,w}$ can be obtained from  $\mathcal{G}_{u,N,w}$ in the next section.

\subsubsection{Estimating the attack feasibility via $k$-anonymity}
In this paper, we study to what extent these countermeasures limit the re-identification problem. We focus on the case of a website $w$ building the Denoised Reconstructed Profiles $\mathcal{R}_{u,N,w}$ for all its users $u\in U_w$.

To estimate the probability of the attack success, we compute the probability of a user $u$ being $k$-anonymized among the audience of a given website. In other words, we study the probability at which at least $k-1$ users $u_i$ expose the same identical Denoised Reconstructed Profiles, i.e., $\mathcal{R}_{u,N,w}=\mathcal{R}_{u_i,N,w} \ \forall i=(1,2,\ldots, l), l>k-1$.

Those users who expose a unique Denoised Reconstructed Profile could be a potential target for re-identification. Conversely, if two or more users in the website audience expose the same Denoised Reconstructed Profile, re-identification would not be possible.

Note that the attacker could leverage other external information than users' topics (e.g., IP address, time-zone, browser, operating system, etc.) to obtain a more specific user fingerprint. This would be equivalent to restricting the population of users the Global Reconstructed Profile signature needs to be compared with, e.g., mounting the attacks to the subset of users having the same external fingerprint. A discussion of this approach is out of the scope of this paper.


\subsection{Eliminating random topic noise}
\label{sec:random-filter}

As introduced by the author of~\cite{thomson2023privacy}, the attacker could identify the random topics $t_{rnd}$ by carrying out a simple statistical test, based on the number of times a topic is exposed by a user. Recall that once a topic has entered the user profile, it remains in it for $E$ consecutive epochs; however, the attacker can easily reconstruct which ``new'' topic  entered the updated Exposed Profile by subtraction with the Exposed Profile of the previous epoch. In a nutshell, the author of~\cite{thomson2023privacy} proposes the attacker keeps only those topics that entered the user profile at least two different times within the window of $N$ epochs.

Here we generalize this attack, with the goal to optimize the threshold below which topics are filtered out for every $N$. Let $p_{rnd}=p/N_{topic}$ be the probability with which a given random topic $t_{rnd}$ replaces a real topic $t^*_e$ during the update of the Exposed Profile $\mathcal{P}_{u,e,w}$. We compute the probability $p^*_{rnd}(f_{rnd},N,p_{rnd})$ according to which such random topic enters the user profile at least $f_{rnd}$ times after $N$ observation epochs. Note that, by construction,  $p^*_{rnd}(f_{rnd},N,p_{rnd})$ represents the probability with which a topic which appears in $\mathcal{G}_{u,N,w}$ exclusively for the effect of random choices being inserted  into $\mathcal{R}_{u,N,w}$.

Being the insertions of a random topic independent and identically distributed at every epoch, $p^*_{rnd}(f_{rnd},N,p_{rnd})$ can be reduced to the probability of observing at least $f_{rnd}$ successes out of $N$ independent Bernoulli trials. 
Therefore it  follows the complementary cumulative of a Binomial distribution:

$$p^*_{rnd}(f_{rnd},N,p_{rnd})= 1-
        \sum_{k=0}^{f_{rnd-1}} 
                \binom{N}{k}p_{rnd}^k(1-p_{rnd})^{N-k}
$$
This allows the attacker to derive the minimal value of  threshold $f_{min}$ that makes $p^*_{rnd}$ smaller than a target probability $p^*_{min}$:

$$f_{min}= \arg \min_f \left(p^*_{rnd}(f,N,p_{rnd})\leq p^*_{min}\right)$$

This strategy works by filtering out the topics that appear less than $f_{min}$ times in $\mathcal{G}_{u,N,w}$. In other words, the attacker includes all the topics that appear at least $f_{min}$ times during the observation period to obtain the Denoised Reconstructed Profile $\mathcal{R}_{u,N,w}$.

Setting $p^*_{min}=10^{-5}$, with the currently proposed $p$ and $N_{topic}$, we obtain $f_{min}=2$ for $N\le30$, and $f_{min}=3$ for $31\leq N<276$.

\subsection{Implications of random topic elimination}

The above filtering algorithm likely removes from $\mathcal{G}_{u,N,w}$ topics whose probability of being selected by the Topic API algorithm is in the order of $p_{rnd}$ (as by construction they will  likely appear less than  $f_{min}$ times in $N$ epochs).

In fact, this is equivalent to filtering out those \textit{real-but-rare topics} the user is interested in, but that she/he seldom visits, so that they appear with very low probability among the top-$z$ topics considered by Topics API algorithm to build the profile history (Step 2). In a nutshell, those rare topics that have a probability of being exposed in the order of  to $p^*_{rnd}$ are at risk of being filtered out.

The resulting Denoised Reconstructed Profile $\mathcal{R}_{u,N,w}$ thus will be different than the actual profile of the target victim. The re-identification attack thus would be possible only if the attacker knows also the probability of the victim exposing a given topic. By ignoring the rare topics, the attacker builds the effective victim profile and matches it with $\mathcal{R}_{u,N,w}$.

Considering instead the re-identification attack carried out  by two websites to match their audiences (or by a third-party $s$ that colludes with website $w_1$ and $w_2)$, both $w_1$ and $w_2$ reconstruct the same Denoised Reconstructed Profile $\mathcal{R}_{u,N,w_1}=\mathcal{R}_{u, N,w_2}$, thus making this attack possible. In this case, the filter does not limit the attack, but we expect it to make the profiles more anonymized since rare topics would be filtered out (making the Denoised Reconstructed Profiles more similar among those of different users).

We expect the proposed denoising algorithm to be very effective in filtering out random topics. As discussed above, the algorithm by design would also filter out true-but-unpopular topics for every user.
\section{Dataset}
\label{sec:dataset}

To simulate the Topic API algorithm in a realistic environment, we rely on a dataset of actual browsing histories collected from a population of users that joined a Personal Information Management System (PIMS).

\subsection{Data collection methodology}
In the context of the PIMCity project\footnote{\url{https://www.pimcity-h2020.eu/}, accessed on \today}, we designed, implemented, and deployed a fully-fledged online PIMS called EasyPIMS and opened it for experimentation~\cite{jha2022pims}. Using EasyPIMS, a user has the possibility to upload their personal information and fully control which data to share and for what purpose. A simple web interface allows the user to provide fine-grained consent for sharing the data with data buyers and eventually to monetise their data in a marketplace. Among various types of data, the platform allows users to share their browsing history by installing a browser plugin for Google Chrome or Microsoft Edge on their PC running any operating system. Such plugin records all \textit{intentionally visited webpages} and stores them in a central repository. During the test of our PIMS, we recruited $3,369$ volunteers who had the possibility of using the platform for four months in 2022. Out of them $928$ installed the plugin. To join the PIMS, there was no restriction on the geographic area, and users belong to $35$ different countries in Europe, Asia, and America. Considering the demographic information of the population, $478$ are male, $226$ are female and $224$ did not declare their gender. The age ranges from $18$ to $72$ years, the average being 33.

In this paper, we leverage the actual browsing histories of EasyPIMS users that explicitly provided their consent for research purposes to the usage of their browsing history and any personal data we use. $613$ gave such permissions. Among those, we restrict the population to those users that actively used the platform. Since the Topics API operates on a weekly basis, we consider a user to be active in a given week if they visited at least $10$ webpages. In total, we obtain $268$ users that result active in at least one week. We use the sequence of websites visited by these users for our study.

\paragraph*{Ethical Aspects}
Our data collection process is compliant with ethical principles and EU privacy regulations. EasyPIMS was part of a European Project involving 12 partners and the European Commission has approved all the data collection and processing procedures. Users voluntarily participated, were informed, explicitly opted-in via the PIMS web interface, and were rewarded by sweepstakes. We only use data of users who explicitly provided their consent for the specific purpose of research, which was not the default choice of the platform. Moreover, data processing has been carried out in an anonymous fashion using a secure computing infrastructure running up-to-date software and with restricted physical access to authorized personnel. During data processing, we only process data regarding browsing histories, neglecting all other attributes, such as name, gender, or geographic location.


\subsection{Characterization of users' activity}

\begin{figure}[t]
    \centering
    \begin{subfigure}[t]{0.25\textwidth}
        \includegraphics[width=\columnwidth]{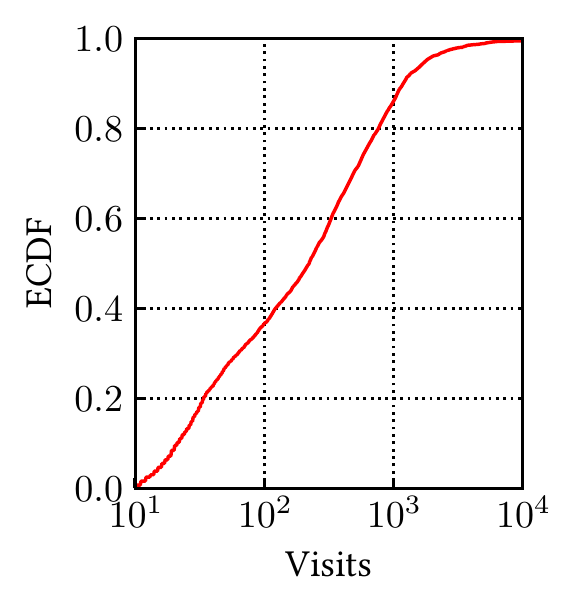}
        \caption{Webpage visits}
        \label{fig:dataset_visits}
    \end{subfigure}
    \begin{subfigure}[t]{0.2135\textwidth}
        \includegraphics[width=\columnwidth]{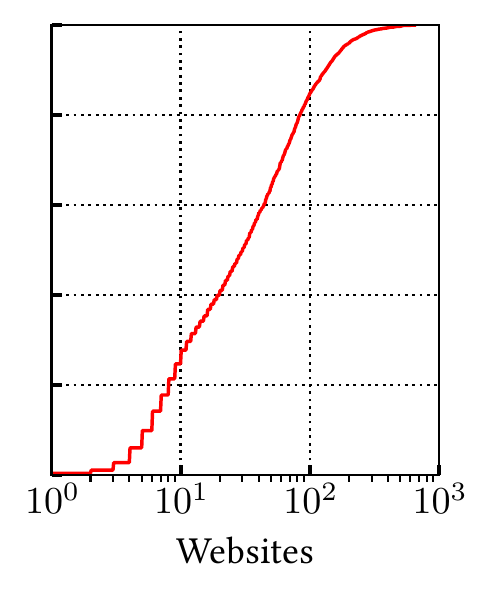}
        \caption{Website visits}
        \label{fig:dataset_websites}
    \end{subfigure}
    \caption{Distribution of the number of visits for webpages and websites per user per week.}
    \label{fig:dataset}
\end{figure}

In total, our dataset includes $2,813,283$ webpage visits to $50,976$ different websites.
The number of visits per user per week varies significantly, with some users that used the platform for a few weeks and others for the whole four-month experimental period.

Some users even installed the plugin on multiple browsers and devices (e.g., desktop and laptop PC), increasing the amount of data collected in their accounts. In detail, we characterize the different usage patterns in Figure~\ref{fig:dataset}. We show the Empirical Cumulative Distribution Function (ECDF) of the number of page and website visits each user recorded each week in Figure~\ref{fig:dataset_visits} and Figure~\ref{fig:dataset_websites}, respectively. We observe a large variability. In the median, active users access $222$ web pages each week, with 26.1\% of users that visit less than $50$ pages; conversely, 14\% of the users visit more than $1,000$ pages. The most active users have accessed about 10,000 pages in a week.

Similar considerations hold when we focus on the number of unique websites a user visits in a week in Figure~\ref{fig:dataset_websites}. On the median, active users access $30$ different websites in a week, while the $25^{th}$ and $75^{th}$ percentiles of the distribution are $10$ and $71$ websites, respectively. The most active users access more than 500 websites in a week.
 
Overall, we believe these figures reflect the natural variability of users. Despite being limited, our dataset includes a real population of users browsing the web, with different interests, backgrounds, nationalities, etc. Unfortunately, we cannot advocate our dataset is representative of general human behaviour and we do not exclude it may be biased in some direction such as gender or education. In the following, we use it to study the impact of the Topic API algorithm to avoid an attacker to mount a re-identification attack. 

\subsection{Characterization of topic statistics}

Using the current implementation of the Topic API ML model Google opened since Chrome 101, for each of the 50,976 websites $w$ in our dataset, we extract the corresponding topic $t$ the API returns. We obtain 250 topics visited at least once by users in our dataset. In the following, we report the characterization of the topic visits.

\begin{figure}
\centering
    \begin{subfigure}[t]{0.227\textwidth}
        \includegraphics[width=\columnwidth]{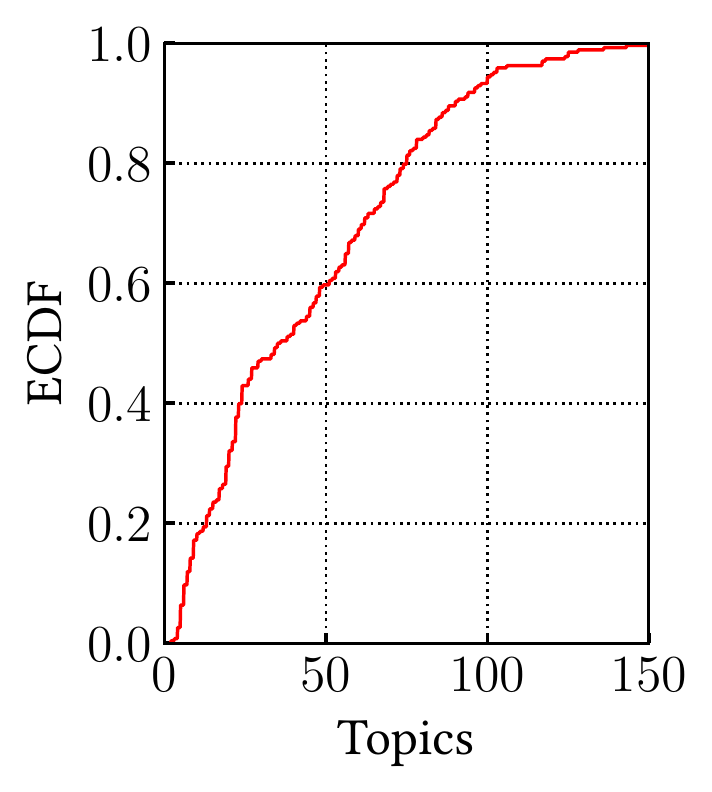}
        \caption{Number of topics per user on the entire observation period.}
        \label{fig:topics_per_user}
    \end{subfigure}
    \begin{subfigure}[t]{0.230\textwidth}
        \includegraphics[width=\columnwidth]{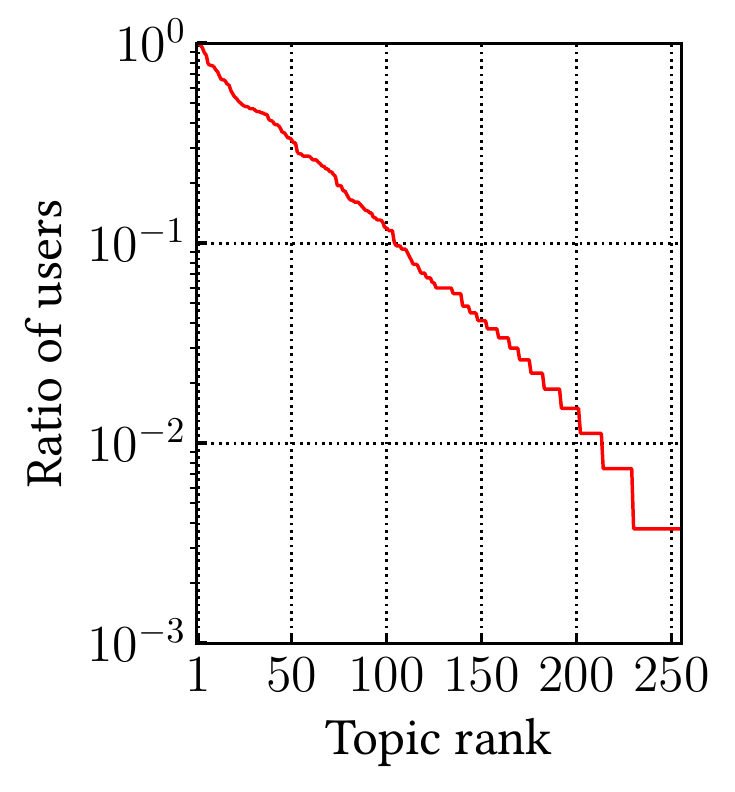}
        \caption{Ratio of users visiting a topic. Topics in decreasing popularity.}
        \label{fig:user_per_topic}
    \end{subfigure}
    \begin{subfigure}[t]{0.230\textwidth}
        \includegraphics[width=\columnwidth]{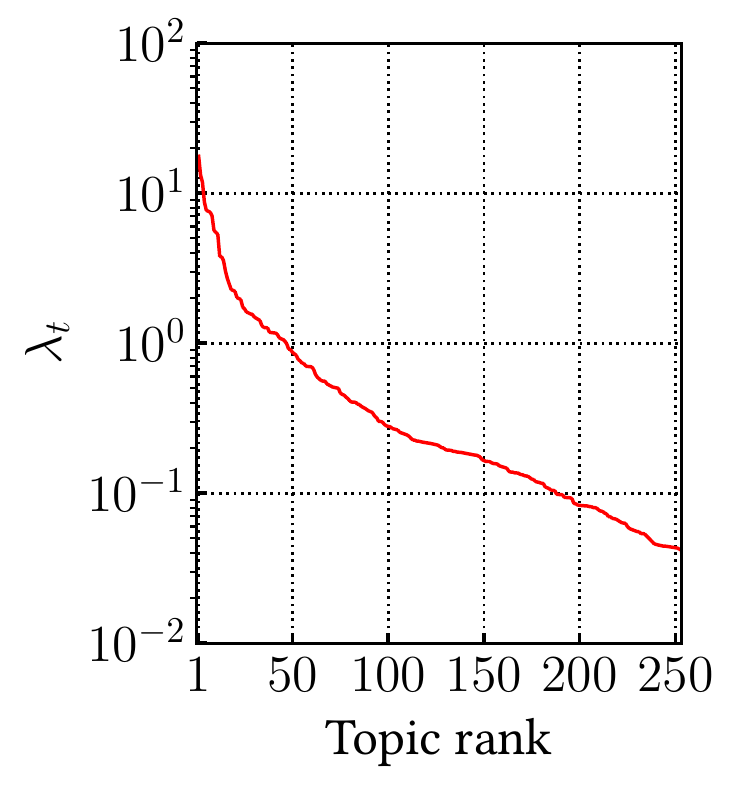}
        \caption{Average rate of visits per topic. Topics in decreasing rate.}
        \label{fig:topic_rate}
    \end{subfigure}
    \caption{Characterization of topic visits.}
    \label{fig:topics_marginals}
\end{figure}

Focus first on the number of unique topics each user visited at least once during the entire experimentation. This is useful to understand how complicated (and unique) could be a Profile $\mathcal{P}_{u,e}$. We report the ECDF in Figure~\ref{fig:topics_per_user}. The distribution is quite spread: in the median users visit 36 topics, with the most diverse users visiting more than 150 topics. Conversely, a handful of users visit less than 5 topics. Not reported here for the sake of brevity, the median number of topics each user visits per week is 17, with a maximum of about 70. Only less than 10\% of users visit less than 5 topics in some weeks.

Figure~\ref{fig:user_per_topic} reports the ratio of users visiting a given topic. We sort topics by their popularity in decreasing order. The ranking follows a clear power-law distribution (notice the log-log scale), as typically happens with popularity distributions in web measurements~\cite{adamic2000power}. The top-5 topics are Search Engines, News, Arts \& Entertainment, Internet \& Telecom, and Business \& Industrial. The most popular topic is visited by 99,3\% of users, while up to 100 (200) topics are visited by at least 10\% (1\%) of the users. 

At last, we show the average rate of visits per topic in Figure~\ref{fig:topic_rate}. We compute first the rate of visits of user $u$ to topic $t$ $\lambda_{u,t} = \sum_e f'_{t,u,e}/ T$, being $T$ the total activity time (discretized by weeks) of user $u$ in the whole observation window.
Then, we compute the average rate of visits among the subset $U_{|t}$ of users that visited the topic $t$ as
\begin{equation}
\lambda_{t}= \sum_{u\in U_{|t}} \frac{\lambda_{u,t}} {|U_{|t}|}
\label{eq:rate}
\end{equation}
Notice a topic that is globally unpopular can still sizeably appear in the Profile of those few users frequently visiting such topic. In fact, the construction of the topic history $\mathcal{T}_{u,e}$ depends on the rate of visits $\lambda_{u,t}$ the user $u$ has for the topics $t$ she/he is interested in during the $e$-th epoch. Our dataset allows us to estimate $\lambda_{u,t}$ for all users.

In the following, we present three models that allow us to generate some possible realistic population $U$ and to study the $k$-anonymity properties of the resulting set of Denoised Reconstructed Profiles $\{\mathcal{R}_{u,N,w}\}$ the website $w$ obtains for such population $U$ after $N$ observation epochs.
\section{Scenarios and User models }
\label{sec:models}

We describe the process we follow to simulate a malicious website that mounts a re-identification attack on the population of its visitors. We create and study the system's behaviour considering different sets of \textit{personas}, i.e., artificial visitors whose statistical parameters are derived from the distribution of the real users in our dataset.

\subsection{Threat model scenario}

We consider a website $w$ that observes and tracks a population of personas $U$ that visit it. The website uses first-party cookies to build the Denoised Reconstructed Profiles $\mathcal{R}_{u,N,w}$ for each persona in $U$. Those personas whose $\mathcal{R}_{u,N,w}$ results unique among $U$ are potentially subject to the re-identification attack.

For simplicity, we assume that at each epoch $e$, all personas visit the website $w$. At each visit, $w$ asks the Exposed Profile $\mathcal{P}_{u,e,w}$ invoking the Topics API. This allows $w$ to discover a new topic at each epoch.\footnote{If the persona does not visit the website for some epochs, the attacker could still recover up to $E$ past topics for every single visit. $N$ can then be considered the number of effective epochs the victims visit $w$ and results in a lower bound to the number of disclosed topics.} After $N$ epochs, $w$ eliminates those topics that have been exposed less than $f_{min}$ times and obtains the Denoised Reconstructed Profile $\mathcal{R}_{u,N,w}$ for all users $u\in U$.

To assess the vulnerability to re-identification attack, we compute the $k$-anonymity properties of the set $\{\mathcal{R}_{u,N,w}\}$ of Denoised Reconstructed Profiles.

\subsection{Population models}
\label{sec:population-models}
We consider three models for the generation of $U$.
The first model follows a mere trace-driven approach that aims at replicating the browsing behaviour of the real users in our dataset. The other models allow us to generate an artificial population $U$ of any desired size $|U|$: the second model obtains personas with the same first-order statistical properties of the users in the trace; the third model consists in a combination of the visiting rates by the users in our dataset.

\vspace{2mm} \noindent
{\bf Real Users:}
In this first model, we consider each of the $268$ users in the dataset. Each user is characterized as a list of visit rates $\lambda_{u,t}$ for all $t=1,2,\ldots,n_{topic}$. $\lambda_{u,t}$ is calculated by averaging the occurrences $f'_{t,u,e}$ along the period in which the user $u$ has been active in our collection system. $\lambda_{u,t}=0$ if that user never visited topic $t$.  Then, for each user $u$, we simulate users' topic visits,
which are assumed to follow independent homogeneous Poisson processes, to build the Profile history $\mathcal{P}_{u,e}$  over epochs $e=(1,2,\ldots,N)$. At each epoch, the website $w$ discovers one topic for every user, and after $N$ epochs it obtains the set $\{\mathcal{R}_{u,N,w}\}$ of Denoised Reconstructed Profiles. While accurate and real, this model is limited to the $268$ users in our dataset.~\footnote{The empirical observation period is unfortunately not long enough to use directly $f'_{t,u,e}$ as input to build the profile history.}

\vspace{2mm} \noindent
{\bf I.I.D. Personas:}
We create a population of i.i.d. personas obeying to the same marginal statistics as the set of real users from our dataset. In detail, we leverage (i) the marginal ECDF of the number of topics per user (Figure~\ref{fig:topics_per_user}), (ii) the marginal empirical distribution of the topic popularity (Figure~\ref{fig:user_per_topic}),  and (iii) the  average empirical rate of visits for each topic $\lambda_{t}$ (Figure~\ref{fig:topic_rate}). In such a way, we can create a population of any size $|U|$ that shares the same first-order statistical properties as the population of our dataset. We adopt the inverse transform sampling method~\cite{devroye1986sample} for the generation of the random variable that follows a known ECDF. In detail, we generate a persona $u$ according to a three-step process:
\begin{enumerate}
    \item We extract the number of topics $c_u$ the persona is interested in from the empirical marginal distribution of the number of topics per user (Figure~\ref{fig:topics_per_user}).
    \item We choose the set of the topics $C_u=\{t_i\},\ i=1,2,\ldots,c_u$ by extracting with no repetitions $c_u$ topics from a normalized version of the empirical  distribution of the topic popularity (Figure~\ref{fig:user_per_topic}).
    \item For each $t\in C_u$, we assign an effective visit rate  $\lambda_{t}$ from Equation~\ref{eq:rate}, which equals the average empirical visiting rate (Figure~\ref{fig:topic_rate}). 
\end{enumerate}
Notice that in step 2 we select each topic essentially independently (just disregarding possible repetitions). This breaks existing correlations among topics and may appear in part  unrealistic. In fact, it is known that real users show highly-correlated interests which reflects in highly-correlated topics~\cite{viswanath2014towards}. The resulting personas in $U$ have instead all the same statistical properties making the probability of having similar profiles high, increasing the probability of being $k$-anonymous. As such this model is a rather pessimistic scenario for the attacker.

\vspace{2mm} \noindent
{\bf Crossover Personas:}
We generate each persona $u$ according to the biologically-inspired crossover procedure during the generation of offspring. We start the process from the population $U^*$ of Real Users. We then randomly select two parent individuals $p_0$ and $p_1$ from $U^*$ and generate a new persona $u$. It inherits part of the genome (i.e., visit rates to topics) from $p_0$ and part from $p_1$. For this, we generate a binary mask and assign the rate of $p_0$ ($p_1$) if the corresponding bit is true (false). In this third model, the correlation of the appearance of topics is stronger than in the previous case. For this, we expect this scenario to be optimistic for the attacker since the uniqueness of personas is boosted by making them more heterogeneous and easier to re-identify.

\subsection{Simulation of visits and profile creation}

Given the population $U$, we assume each persona $u$ visits topic $t$ according to a homogeneous Poisson process with the assigned rate $\lambda_{u,t}$. At each epoch $e$, for each topic $t$ and persona $u$, we thus extract a Poisson distributed random variable that represents the number of visits user $u$ performs to $t$. This allows us to obtain the topic history $\mathcal{T}_{u,e}$, and from it the Profile history $\mathcal{P}_{u,e}$ which contains only the top-$z$ topics (Step 2 of Topic API algorithm). Next, we generate the Exposed Profile $\mathcal{P}_{u,e,w}$, possibly offering $w$ a random topic instead of a real top topic (Step 3).

By repeating the periodic profile update procedure at the beginning of each epoch $e+1$, we simulate the process for $N$ epochs so that at the end $w$ compiles the Denoised Reconstructed Profile $\mathcal{R}_{u,N,w}$ for each persona $u\in U$.
\section{Results}
\label{sec:results}

In this section, we illustrate the results of our study. We first quantify the probability of a persona to be 2-anonymized after different observation epochs $N$. Then, we evaluate the impact of the population size $|U|$. Finally, we discuss to what extent the injection of random topics and the attacker's filtering strategy impact the 2-anonymity probability.

In the following, where not expressly stated, we set $p^*_{min}=10^{-5}$ and consider the Google suggested values for the Topic API parameters ($z=5,\ E=3,\ p=0.05, \Delta T=1$~week). We repeat each experiment 10 times and report the average performance.

\subsection{Probability of being 2-anonymous}

\begin{figure}
    \centering
    \includegraphics[width=\columnwidth]{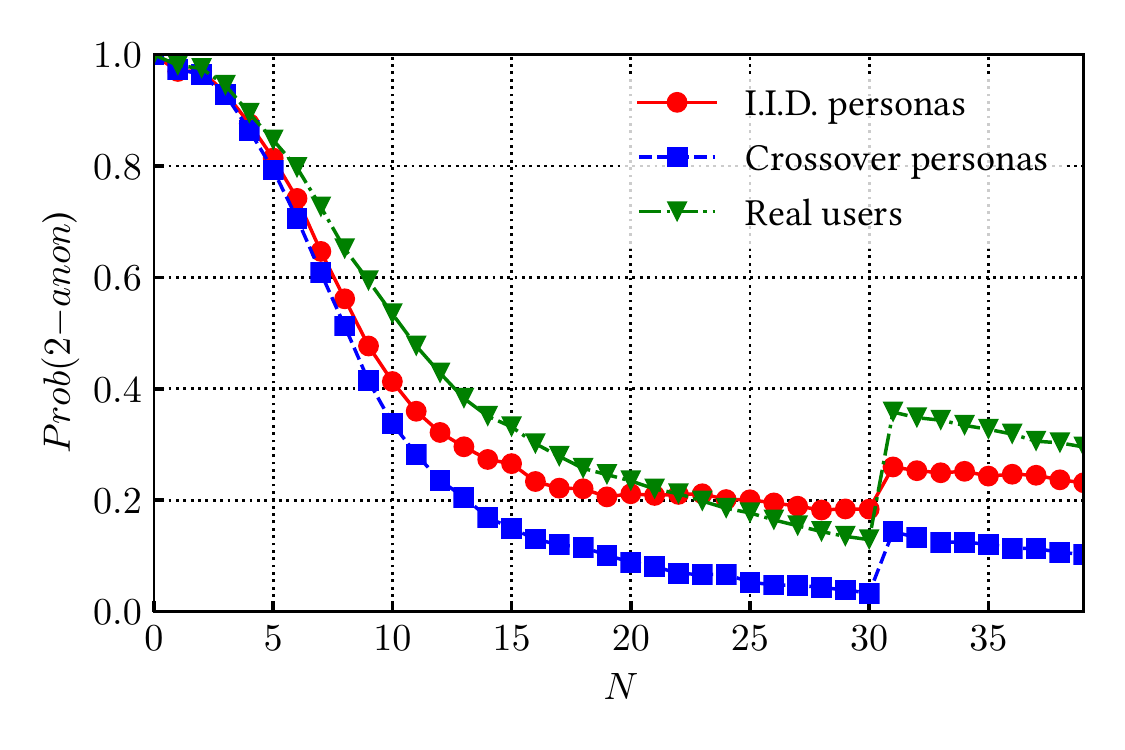}
    \caption{The probability of a user being 2-anonymized with different audiences of $|U|=268$ users/personas.
    Without the Topics API $Prob(2-anon)\approx 0.17$ for real users.}
    \label{fig:compare_methods}
\end{figure}

We start our analysis by showing the effectiveness of a re-identification attack mounted on the profiles
$\{\mathcal{R}_{u,N,w}\}$ which $w$ builds using the Topics API. As described in Section~\ref{sec:threat}, we consider the attack on user $u$ not feasible whenever $u$ is $k=2$ anonymized. Therefore, we measure the probability \Panon to be 2-anonymous among personas in $U$. 

Here, we focus on a population of users/personas generated according to the three models described in Section~\ref{sec:models}. In Figure~\ref{fig:compare_methods}, we show the probability of such persona to be $2$-anonymized after $N$ observation epochs. As the simulations with real users are limited to $268$ individuals by design, we carry out the study with 268 personas for I.I.D. and Crossover models for a fair comparison. 

For the first epochs, all Denoised Reconstructed Profiles result 2-anonymous because the attacker obtains few topics in which each persona is interested. In fact, for $N=1$, all profiles $\mathcal{R}_{u,1,e}$ result empty because, by construction, the denoising algorithm discards the only topic $u$ offers to $w$ (no topic can appear more than $f_{min}=2$). As epochs pass, the attacker accumulates topics and starts creating unique Denoised Reconstructed Profiles. The \Panon starts decreasing until reaching a first asymptote: the attacker is able to reconstruct rich profiles and only $3$ to $20\%$ of personas results $2$-anonymous for all three population models. At epoch $31$, suddenly the \Panon increases significantly for all models. This is the result of the increase of $f_{min}$ from 2 to 3 to match the random topics filter target  (probability $p^*_{min}=10^{-5}$).
As a consequence, the attacker starts to filter out also actual-but-rare topics, building more homogeneous Denoised Reconstructed Profiles. This effect appears every time the attacker has to increase $f_{min}$. Apparently, setting $N=[25,30]$ maximises the attack success probability. In the following, we choose $N=25$.


Compare now the three models to generate the audience. In the initial phase, we observe little differences between them, and the fraction of 2-anonymous personas rapidly decreases for $N=[5,15]$.
Note that the Crossover personas (blue dashed line) have a lower \Panon compared to other methods. This is expected because the model generates more specific profiles by design. Conversely, I.I.D personas (solid red line) apparently settle to a $2$-anonymization probability of $\approx0.2$. This is due to the homogenous profiles the model generates so that the probability of being 2-anonymous increases.

We now compare these figures to the scenario without the Topic API algorithm. We assume the attacker builds users' profiles directly on the set of topics $\mathcal{T}_{u,e}$. This is similar to what happens today, where web trackers are free to directly build the topic list from users, with no filters. In this case, we compute \Panon on the set $\mathcal{T}_{u,e}$. It results in \Panon=~0.17 for our real user data. Comparing this result with the curves (and with the Real users in particular) in Figure~\ref{fig:compare_methods}, we observe that Topics API offers limited additional protection, provided the attacker can observe users for a sufficient time.

In a nutshell, the Topics API allows an attacker to still build specific user profiles so that they can be uniquely identified with high probability. Yet it will need $25$ to $30$ epochs to reconstruct the profiles. The $2$-anonymization resulting probability is very similar to \Panon=~0.17 a web tracker would obtain without the Topic API filter.

\subsection{Impact of population size}

\begin{figure}
    \centering
    \includegraphics[width=\columnwidth]{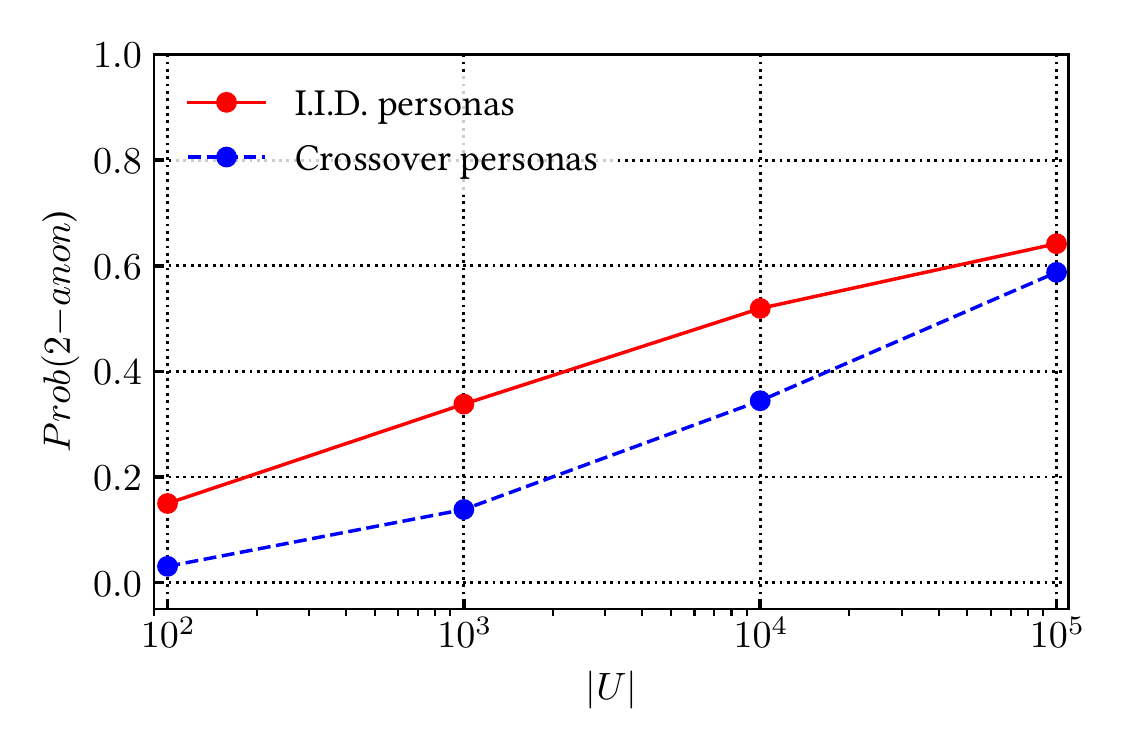}
    \caption{Probability of being 2-anonymized with different numbers of personas.}
    \label{fig:nusers}
\end{figure}

We now quantify the impact of the population size on the probability of a persona being 2-anonymized. Indeed, by increasing a website's audience, we increase the probability of finding other personas who match the same profile. We measure this effect in Figure~\ref{fig:nusers} where we show how \Panon varies with the population size $|U|$.

As expected, larger audiences increase the chance to be 2-anonymized. 
Yet, the probability increases roughly logarithmically with the cardinality of $U$ -- notice the log scale on the $x$-axis. Thus, even in the case of a population of $100,000$ personas, we still observe $35-42\%$ of them being non-2-anonymous. Again, the homogeneous I.I.D. personas model constitutes a worst-case for the attacker.


In a nutshell, the Topic API algorithm improves the $2$ anonymity. Still, a relevant fraction of users remains still non-2-anonymous even in the case of large populations. 
Notice that the attacker can combine other information to reduce the population size, e.g., by considering users with the same region, timezone, browser, operating system, etc. This would partition $U$ into  subsets of smaller cardinality, where the probability of not being 2-anonymous significantly increases.


\subsection{Different $k$-anon levels}

\begin{figure}
    \centering
    \includegraphics[width=\columnwidth]{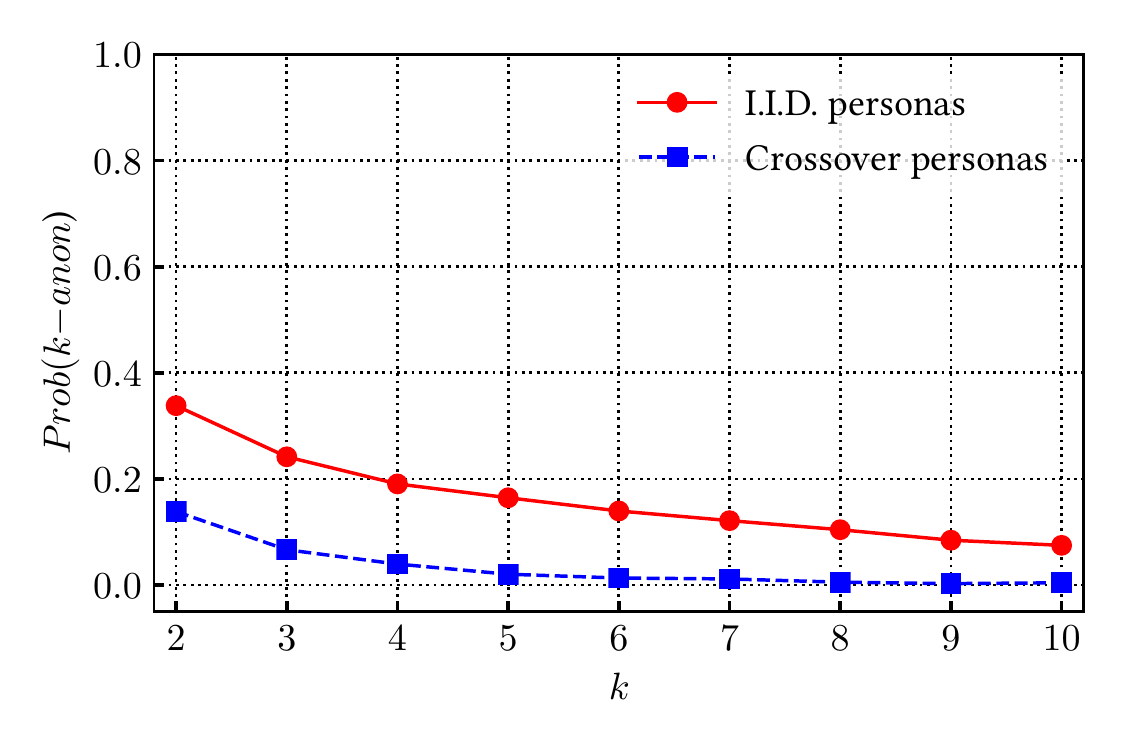}
    \caption{Probability of being $k$-anonymized with Topic API active.}
    \label{fig:kanons}
\end{figure}

We generalize the above discussion by quantifying the $k$-anonymity offered by the Topics API. Generally speaking, we consider a persona \emph{anonymized} if their Denoised Reconstructed Profile $\mathcal{R}_{u,N,w}$ is not unique -- i.e., it is $2$-anonymized. Yet, if there are only two personas sharing the same profile, with a probability of $0.5$ one could identify the correct persona.
Here, we observe the probability the target profile is shared by $k-1$ other users, i.e., of being $k$-anonymous. Intuitively, a higher $k$ offers stronger privacy guarantees.

Figure~\ref{fig:kanons} shows $Prob(k\text{-}anon)$ for different values of $k$. Here we consider a population of $|U|=1,000$ and $N=25$. We show curves for I.I.D. and Crossover personas, with the former being a worst-case for the attacker, the latter being more realistic in practice.

As expected, with increasing values of $k$, the fraction of $k$-anonymized personas decreases. Interestingly,  observe how the $k$-anonymity probability decreases faster for Crossover personas than for I.I.D. personas. For instance, in the latter case, a persona has $\approx 10\%$ of possibilities to share a profile with $10$ or more other individuals, while the probability is negligible ($0.002$) for Crossover personas.

In a nutshell, the more heterogeneous the users, the harder is to be $k$-anonymous and the Topic API algorithm does not prevent an attacker to mount a re-identification attack.

\subsection{Impact of random topic protection}

\begin{figure}
    \centering
    \includegraphics[width=\columnwidth]{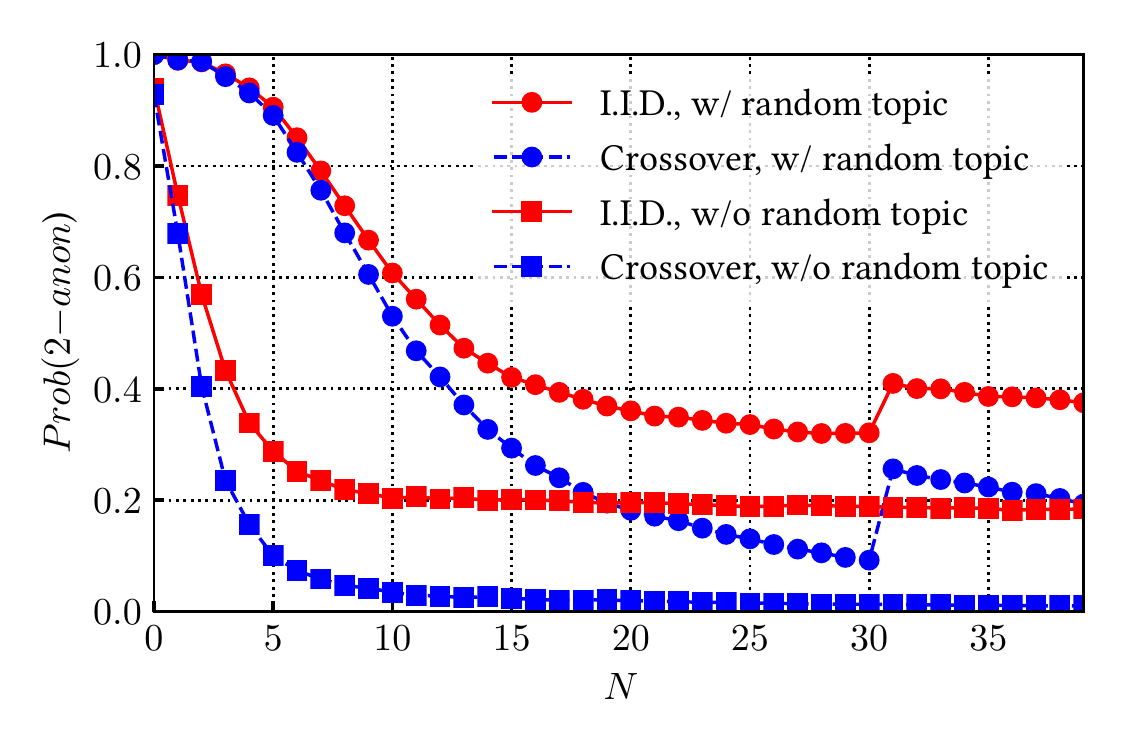}
    \caption{Probability of being 2-anonymized with and without the replacement with random topic.}
    \label{fig:topic_vs_non_topic}
\end{figure}

\begin{figure}
    \centering
    \includegraphics[width=\columnwidth]{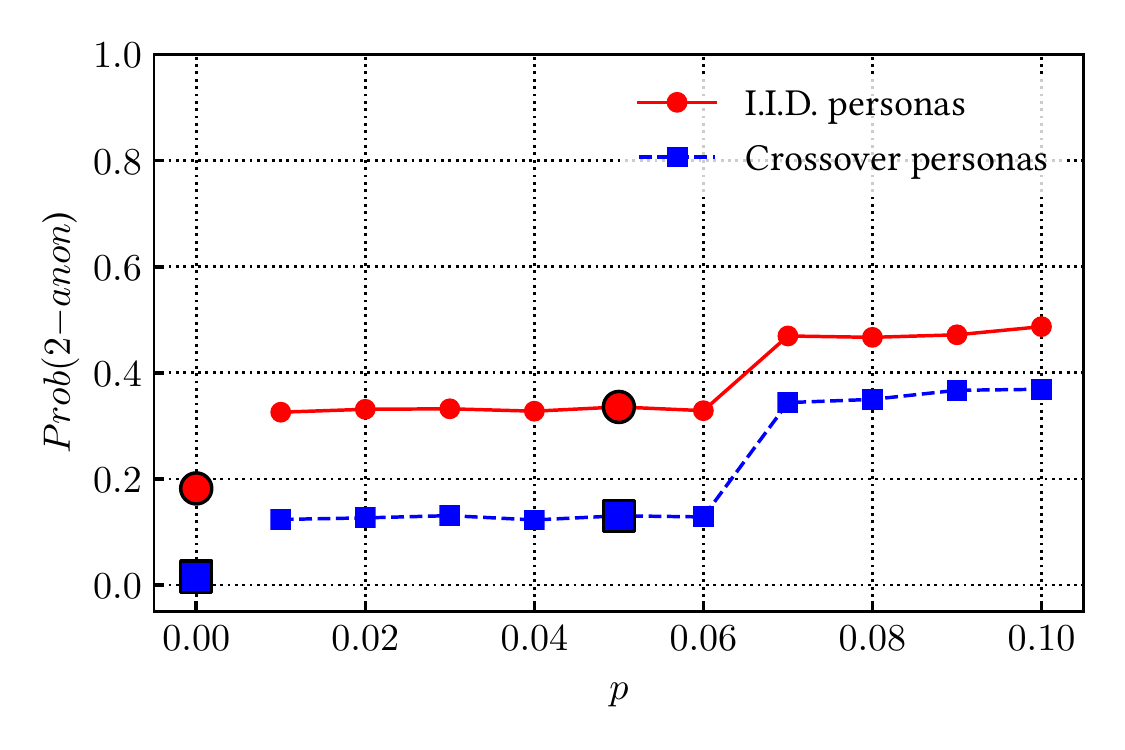}
    \caption{Probability of a user being 2-anonymized, with different levels of $p$. We highlight the points at $p=0$ (no random topics) and the points at $p=0.5$ (the default value proposed for Topics API.}
    \label{fig:ps}
\end{figure}

We finally study the implications of replacing a real topic with a random topic on the privacy properties offered by the Topics API algorithm. 

\subsubsection{Benefit of topic replacement}

Here, we quantify the benefits offered by random topic replacement on \Panon. To this end, we compare  scenarios where Topic API algorithm disables ($p=0$) or enables ($p=0.05$) the random topic replacement policies. When disabled, the attacker does not adopt any filtering strategy.

In Figure~\ref{fig:topic_vs_non_topic} we compare results considering  $U=1,000$ personas, and for different  observation epochs $N$. The role of random topic appears clear since the first epochs: When not active (squared dots), the probability of anonymization quickly decreases with the observation epochs so that the attacker can build a unique profile already after 10 epochs. With I.I.D. personas, 20\% of the population exhibits the same profile. For Crossover personas, in practice, each profile results unique (\Panon=0.02).

Activating the random topic replacement increases the probability of being 2-anonymous to a remarkable extent. \Panon never falls below $37\%$ ($8\%$) for I.I.D. (Crossover) personas. This is caused by the denoise filter that the attacker must activate to filter the random topics. In fact, it causes  rarely exposed topics to be filtered as well, which leads to an increase in the homogeneity of Denoised Reconstructed Profiles, and ultimately, to an increase of 2-anonymization probability. Again, the step at $N=31$ reflects the need to set $f_{min}=3$, increasing further the homogeneity of the Denoised Reconstructed Profile. Note that when random topics are disabled, no filter is in place and $f_{min}$ is always $0$.

In a nutshell, without random topic replacement, the re-identification attack would turn quite simple. Inserting random topics helps in increasing the 2-anonymity probability. Yet, the probability of having a unique Denoised Reconstructed Profile is quite high even when we consider optimistically similar I.I.D. personas.

\subsubsection{Impact of the choice of $p$}

At last, we quantify the role of the probability of exposing a random topic $p$. In Figure~\ref{fig:ps} we show how \Panon varies with different values of $p$. We consider a population of $1,000$ users observed for $N=25$ epochs. $p=0.05$ corresponds to the suggested value; $p=0$ means disabling the random topic replacement and the denoising algorithm.

Considering the trend, the figure shows that, overall, $p$ has a somewhat quantized impact. Considering I.I.D. personas (solid red curve), $p=0$ leads to $Prob(2-anon) = 0.19$. This probability suddenly increases to $\approx 37\%$ for all values of $p \in [0.01, 0.06]$, for which $f_{min}=2$ suffice to guarantee to filter the random topics with $p^*_{min}=10^{-5}$. Then, when $p \ge 0.07$, \Panon jumps to $\approx 45\%$. This is because $f_{min}$ needs to be set to 3 to filter the random topics. In turn, this causes the filtering of additional actual-but-rare topics, and thus to more homogeneous Denoised Reconstructed Profiles.
The shape is similar for Crossover personas (blue dashed curve) but with lower \Panon values.

In a nutshell, while injecting random topics is beneficial to increase the 2-anonymity properties of the set of Denoised Reconstructed Profiles, the noise filtering algorithm is robust to a large range of $p$. At last, notice that turning $p$ too large would reduce the benefits of personalised content as well. Notice that the attacker could also optimize the choice of $p^*_{min}$ to balance the effect of allowing some random topics and removing some actual-but-rare topics. We leave this for future work.

\section{Probability of Re-identification}
\label{sec:attack}

After having explored the impact of parameters on the probability of a user being $k$-anonymized, we now simulate the complete re-identification attack described in Section~\ref{sec:threat}. We consider two websites $w_1$ and $w_2$ with populations $U_1$ and $U_2$, with $|U_1|=|U_2|=999$. By construction, we add the same persona $v$ (the victim) to both $U_1$ and $U_2$. We then evaluate the probability of re-identify $v$ by $w_1$ and $w_2$.
For $v$ being to re-identified, two conditions shall occur:
\begin{itemize}
    \item The Denoised Reconstructed Profile $\mathcal{R}_{v,N,w_i}$ is unique in $U_1$ and $U_2$. 
    \item $w_1$ and $w_2$ reconstruct the same profile for $v$, i.e., $\mathcal{R}_{v,N,w_1}=\mathcal{R}_{v_2,N,w_2}$. Indeed, the Denoised Reconstructed Profiles can differ due to the randomness of the topic selection and denoising processes. 
\end{itemize}
Let
$$P_1=Prob\left(\mathcal{R}_{v,N,w_i} \text{ unique in } U_i\right)= 1-Prob(2\text{-}anon).$$
Therefore $v$ results unique in both populations with probability $P_1^2$.


Let $$P_2=Prob\left(\mathcal{R}_{v,N,w_1} = \mathcal{R}_{v,N,w_2}\right).$$ 
$P_2$ is the probability of correctly re-identifying $v$.
Thus, due to the independence of previous events, the probability of \textit{correct} re-identification, i.e., a True Positive (TP), can be computed as:
$$Prob(correct\ re\text{-}identification)=P^2_1\cdot P_2.$$
Similarly, let
\begin{align*}
\overline{P_2}=& Prob\left(\exists !\, v'\in U_2,\ v'\neq v\, :\; \mathcal{R}_{v,N,w_1} = \mathcal{R}_{v',N,w_2}\right.
            \\
            &\left. \mid \mathcal{R}_{v,N,w_1} \text{ unique in } U_1\right).
\end{align*}
$\overline{P_2}$ is the probability of \textit{incorrect} re-identification given $v$ is unique un $U_1$.

The probability of an incorrect re-identification, i.e., a False Positive (FP), becomes:
$$Prob(incorrect\ re\text{-}identification)=P_1 \cdot \overline{P_2}.$$
In other words, given a match between two unique profiles $v\in U_1$ and $v'\in U_2$, the re-identification is successful and correct, i.e., a TP, if $v'=v$. If instead $v'\neq v$, the re-identification is successful but wrong, i.e., a FP.

In the following, we report the results we obtain by repeating for all 1,000 users in $U_1$ the re-identification test. We consider each persona in $U_1$ as a possible victim $v$ (and each time we replace a user in $U_2$ with $v$).\footnote{This is equivalent to assuming $U_1=U_2$.}
As before, we repeat the experiment 10 times and report the average results. We consider both I.I.D. and Crossover populations and observe the TP and FP rate as they evolve over the observation epochs $N$.
\begin{figure}
    \centering
    \begin{subfigure}[t]{0.9\columnwidth}
        \includegraphics[width=\columnwidth]{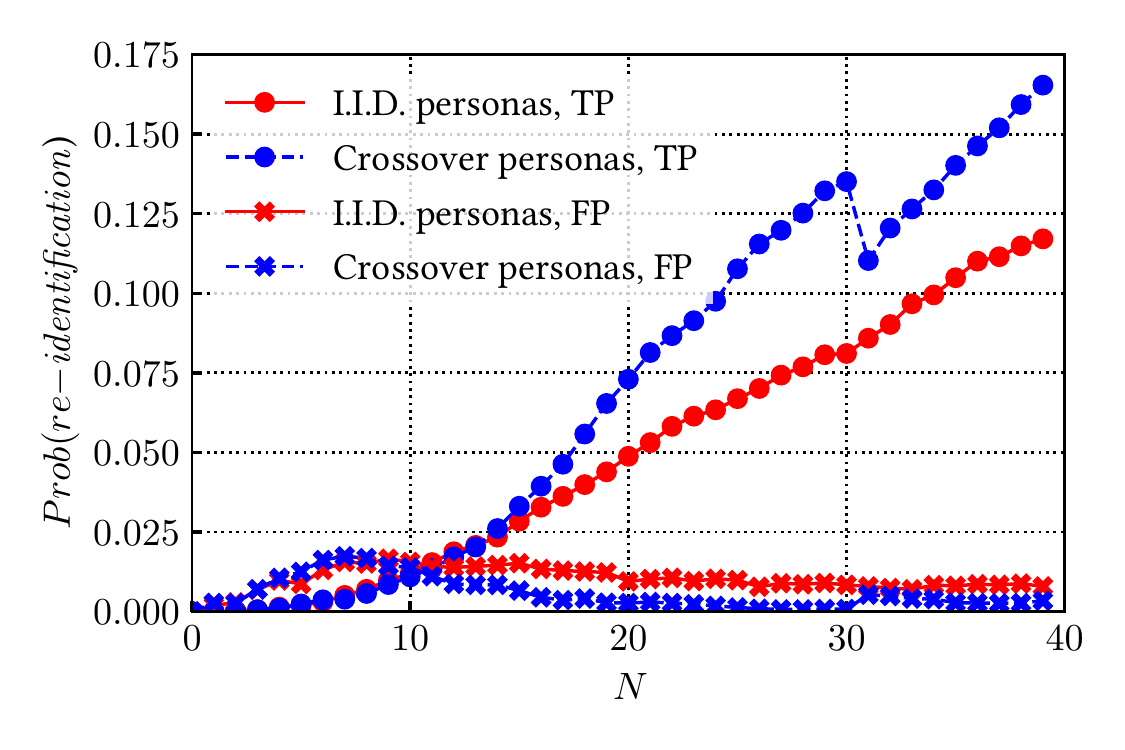}
        \caption{Denoising algorithm in place.}
        \label{fig:attack_filter}
    \end{subfigure}
    \begin{subfigure}[t]{0.9\columnwidth}
        \includegraphics[width=\columnwidth]{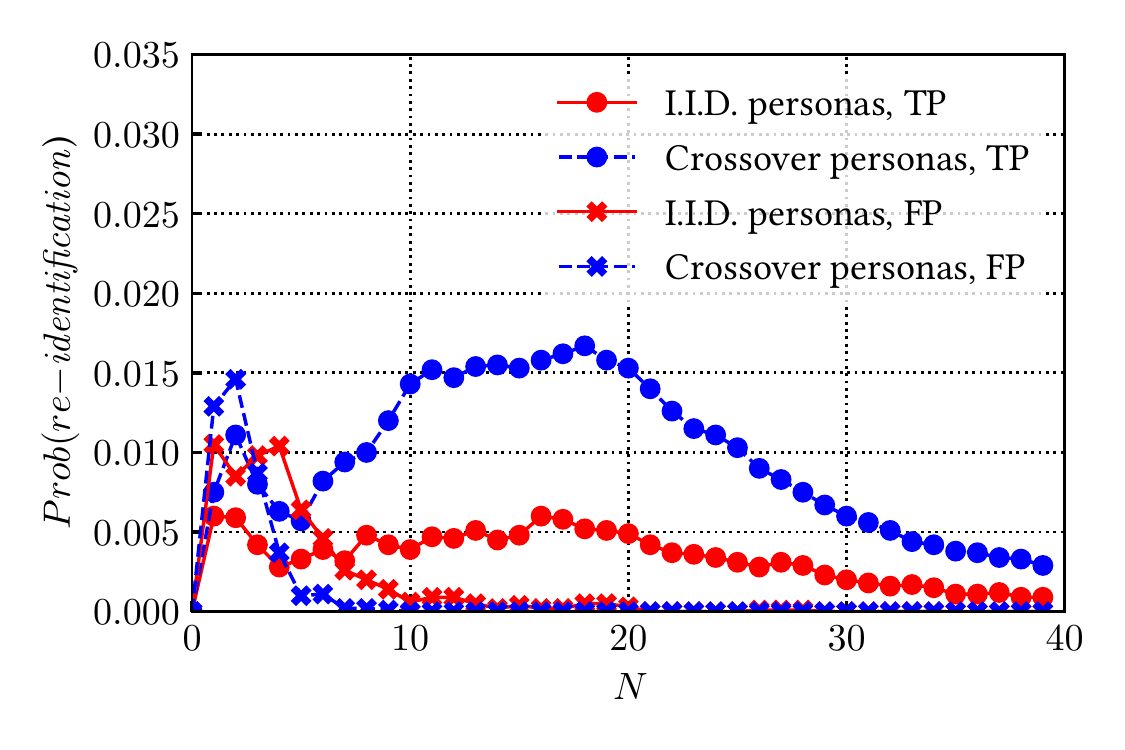}
        \caption{Denoising algorithm not in place. Please note the difference in the y-axis scale.}
        \label{fig:attack_no_filter}
    \end{subfigure}
    
    \caption{Probability of a user being re-identified over two websites with a $1,000$ user audience each.}
    \label{fig:attack}
\end{figure}

We show the results in Figure~\ref{fig:attack_filter}. Focus first on the probability of being correctly re-identified (TP, circled dots). It increases with $N$, i.e., the more the attackers observe the victims, the higher the probability of correctly re-identify them. The scenario with Crossover personas appears more favourable to the attacker since the population of users is  more heterogeneous than the I.I.D. scenario, thus \Panon results small, and the \emph{Prob(correct re-identification)} pretty large. Conversely, in the scenario with I.I.D. personas the construction of unique profiles is more challenging, i.e., $P_1$ is smaller, and thus the re-identification is harder. 
The step-effect at $N=31$ epochs is again due to the automatic increase in $f_{min}$ described in Section~\ref{sec:random-filter} and observed, for instance, in Figure~\ref{fig:compare_methods}. The rise of \Panon causes the drop in \emph{Prob(correct re-identification)}.

Focus now on the probability of incorrect re-identification (FP, crossed dots in Figure~\ref{fig:attack_filter}). During the initial part of the profile reconstruction, the probability of finding the same-but-wrong profile grows. This is because all profiles tend to initially be very similar (see Figure~\ref{fig:topic_vs_non_topic}). As soon as the profiles start to differentiate well, the FP rate decreases. Once more, I.I.D. personas,
being more similar,  lead more easily to FPs  than Crossover personas. In the latter case, the \emph{Prob(incorrect-identification)} is smaller than 0.002 for $N\in[25,30]$, 
and about two orders of magnitude smaller than \emph{Prob(correct re-identification)}.

As a baseline, in Figure~\ref{fig:attack_no_filter} we show the \emph{Prob(re-identification)} when the attacker does not apply the denoising algorithm (hence, comparing Global Reconstructed Profiles $\mathcal{G}_{v,N,w_i}$). The 
\emph{Prob(correct re-identification)} does not reach the 2\%, because the random topics exposed on different websites by users make their profiles more difficult to re-identify. Moreover, the trend over time differs with respect to the one shown in Figure~\ref{fig:attack_filter}: the larger the observation time, the larger the probability that every user would expose some different random topics.

Here, we suppose the attacker looks for an exact match between two user profiles, while it would be possible to use more flexible and effective techniques to link the two sets~\cite{herrmann2013behavior,vassio2017users}.  We leave the evaluation of these alternative approaches for future work.


\section{Related work}
\label{sec:related}

From the dawn of the Web, behavioural advertising has been a pillar of the ecosystem and entailed the collection of personal information through web tracking. This phenomenon has been the subject of several studies that measured its spread~\cite{metwalley2015online,englehardt2016online} or dug into its technical operation~\cite{acar2014web,rizzo2021unveiling,papadogiannakis2021}. The implications of web tracking on users' privacy have become more and more debated by the industry~\cite{rfc7258} and by the research community~\cite{sipior2011online,mayer2012third,estrada2017online}. It also fostered the birth of anti-tracking tools (i.e., the Ad and Tracker Blockers~\cite{pujol2015annoyed}) and encouraged the legislator to issue privacy-related regulations, such as the US CCPA~\cite{ccpa} or the European GDPR~\cite{gdpr}.

Federated Learning of Cohorts (FLoC) has been the first public effort by Google to go beyond the classical web tracking based on third-party cookies~\cite{ravichandran2021evaluation}. In FLoC, users were grouped in cohorts according to the interests inferred by each one's browser. When asking for information about a user visiting a website, third parties were offered the user's cohort, from which they could have information about the user's interests. In the intention of the proposal, FLoC provided an acceptable utility for the advertisers, while hiding the user (and thus, her identity) behind a group of peers~\cite{epasto2021clustering}. However, criticism arose around the easiness for first- and third-party cookies to follow the user over time exploiting the sequence of cohorts to which she belongs to isolate and thus identify her~\cite{rescorla2021technical}. The attack can exploit browser fingerprint to further improve its effectiveness~\cite{berke2022privacy}. FLoC's privacy anonymity properties can be broken in several ways~\cite{turati2022analysing}. As a response to the critics towards FLoC, Google retired the proposal and conceived the Topics API, whose functioning we describe in Section~\ref{sec:topics-api}.

The Topics API exposes users' profiles in terms of topics of interest to the websites and advertising platforms. In this paper, we study to what extent users' profiles can be used by an attacker to re-identify the same individual across time or space. Past works already demonstrated that profiling users based on their browsing activity can present severe risks to the privacy of the users~\cite{estrada2017online}. They can be identified with high probability based on the sequence of visited websites~\cite{olejnik2012why,herrmann2013behavior,vassio2017users}. Mitigation such as partitioned storage has been put in place to limit the risk, but ways to bypass them exist~\cite{randall2022measuring}. Specifically to the Topics API, the same threat we analyze has been already identified by Epasto~\emph{et al.}~\cite{epasto2022measures} from Google. The authors carry out an information theory analysis and conclude that the attack is hardly feasible. In this paper, we go a step further. Our analyses are not limited to an analytical study on profiles' uniqueness but offer a thorough evaluation using real traffic traces and different user models.

While writing this paper, proponents from Google published a new work discussing the privacy implications of the Topics API in~\cite{carey2023measuring}. They define a theoretical framework to determine re-identification risk and test it on Topics API. Differently from us, they do not consider the use of any denoising algorithm.

To the best of our knowledge, Thomson~\cite{thomson2023privacy} from Mozilla has issued the first independent study on the privacy guarantees of Topics API, elaborating on the conclusions by Epasto~\emph{et al.}~\cite{epasto2022measures}. He again used analytical models and raised severe concerns about the offered privacy guarantees. We inspire our strategy for random topic filtering from Thomson~\cite{thomson2023privacy}.

\section{Discussion and Future Work}
\label{sec:conclu}

\paragraph {Summary} The Topics API represents a prominent proposal to replace the current web-tracking solutions based on third-party cookies with a more privacy-friendly approach. In this paper, we have considered the scenario where an attacker carries out a re-identification attack accumulating the topics a website gets via the Topics API to build a unique user profile. 
Our experiments show that such an attack can be successful, provided the attacker observes the victim for enough epochs.

We showed how the replacement of actual topics with random ones is fundamental for limiting the reconstruction of users' profiles. We designed an algorithm to overcome such protection so that the attacker is able to denoise the reconstructed profiles and remove random topics. This makes the denoised reconstructed profiles robust, and the attack possible for a large range of the probability $p$ of a random topic replacement happening. All in all, we showed how the re-identification attack mounted by websites succeeds in about 15-17\% of users, with a negligible probability of a false re-identification.

\paragraph {Limitations} While the attack is possible, several points need to be considered to judge the actual feasibility of such an attack:
\begin{itemize}
    \item The time needed makes it impractical: Given the suggested epoch duration of one week, the attacker needs 20 to 30 weeks (i.e., $\approx 6$ months) to successfully reconstruct the victims' profiles.
    \item During such time, the victim has to visit the attacker's website every week (or at least every $E$ weeks). If this does not happen, the attacker would need even more time to accumulate enough topics the victim is interested in.
    \item The victim's interests may change over time. This is not critical for the re-identification attack mounted by two websites (as they will observe the victim during the same time period). But this may harm the re-identification attack against an \emph{a priori} known victim profile.
    \item The larger the population, the harder the attack. Yet, the attacker may leverage external information to partition the audience and thus increasing the probability of a successful attack.
\end{itemize}

\paragraph{Improvement of Topics API} Being a draft proposal, there is still room to discuss possible improvements to the Topics API. For instance:
\begin{itemize}
    \item Periodically deleting first-party cookies would bring an immediate privacy-related benefit, reducing the number of epochs to build profiles that could be matched across websites. Figure~\ref{fig:attack} shows that deleting the first-party cookies every $N=10$ epochs, for example, would keep the \emph{Prob(correct re-identification)} below 2\% with the current attack setup, with comparable \emph{Prob(incorrect re-identification)}.
    
    \item The default values of $z$, $E$, and $p$ proposed in the draft proposal of Topics API are open to further review. We believe that the code we offer can be used as a tool to investigate how different parameter choices impact the probability of re-identify users, choosing the best combination.
    
    \item The profile of a user $\mathcal{P}_{u,e}$ is currently populated by the $z$-top topics in epoch $e$. It could be worth considering other approaches to build the profile, and better balance the utility of the advertisers and the privacy of the users.
\end{itemize}

\paragraph{Future directions} Our work provides a first study on how re-identification attacks can be carried out against the Topics API, and several angles are still to be explored. In fact, our experiments can be easily extended in several directions.

First, we rely on a dataset collected from the set of volunteers that participated in the EasyPIMS experimentation. As such, we cannot verify the dataset is representative of general human behaviour. The process of gathering such kind of personal data is cumbersome, but a larger and more heterogeneous audience may help in drawing more solid conclusions. In a similar direction, it is interesting to evaluate more diverse population generation approaches, including diverse usage patterns, classes of users, etc. Similarly, the study of the Topics API parameters can be extended to balance the utility and privacy of exposed information.

Other research directions include the extension of the threat model.
The attacker could consider more sophisticated techniques to match the profiles rather than a simplistic exact match on the presence of a topic. For instance, the attacker could leverage the frequencies with which topics are exposed. Or can design some maximum-likelihood algorithm to maximize the correct re-identification probability.
In the literature, several methodologies have been proposed that can be reused for this goal.

Finally, we suppose the attacker has no background knowledge of the victims. As said above, this assumption can be relaxed to study to what extent any additional information on the user (e.g., retrieved through browser fingerprinting techniques) can help the attacker.

The Topics API is a novel proposal by Google, and we believe the research community should further work to understand the implications of its design, as it might become the \emph{de facto} standard for online advertising in the near future.

\begin{acks}
This work was partially supported by project SERICS (PE00000014) under the MUR National Recovery and Resilience Plan funded by the European Union - NextGenerationEU and the project "National Center for HPC, Big Data and Quantum Computing", CN00000013 (Bando M42C – Investimento 1.4 – Avviso Centri Nazionali” – D.D. n. 3138 of 16.12.2021, funded with MUR Decree n. 1031 of 17.06.2022). 
\end{acks}

\bibliographystyle{ACM-Reference-Format}
\bibliography{biblio}

\end{document}